\documentclass[10pt,preprintnumbers,showpacs,amsmath,amssymb,floatfix,twocolumn,prl]{revtex4}   
\usepackage{graphicx}
\usepackage{dcolumn}
\usepackage{bm}
\usepackage{longtable}
\usepackage{graphics}
\usepackage{amssymb}
\usepackage{amsmath}
\usepackage{xspace}
\usepackage{epsfig}
\usepackage[usenames]{color}
\newcommand {\PF}{(TMTSF)$_2$\-PF$_6$ }
\newcommand {\ClO}{(TMTSF)$_2$\-ClO$_4$ }
\newcommand {\ClOn}{(TMTSF)$_2$\-ClO$_4$}

\newcommand {\PFn}{(TMTSF)$_2$\-PF$_6$}
\newcommand {\SRO}{Sr$_2$RuO$_4$ }

\newcommand {\etal}{{\it et al}. }
\newcommand {\etaln}{{\it et al}.}
\newcommand {\etalc}{{\it et al}., }
\begin{document}
\title{A phenomenological model of the superconducting state of the Bechgaard salts}
\author{B. J. Powell}
\email{powell@physics.uq.edu.au} \affiliation{Department of Physics,
University of Queensland, Brisbane, Queensland 4072, Australia}


\begin{abstract}
We present a group theoretical analysis of the superconducting state
of the Bechgaard salts, e.g., \PF or \ClOn. We show that there are
eight symmetry distinct superconducting states. Of these only the
(fully gapped, even frequency, $p$-wave, triplet) `polar state' is
consistent with the full range of the experiments on the Bechgaard
salts. The gap of the polar state is ${\bf d}({\bf
k})\propto(\psi_{u{\bf k}},0,0)$, where $\psi_{u{\bf k}}$ may be any
odd parity function
that is translationally invariant. 
\end{abstract}

\maketitle

One of central challenges facing theoretical physics is the full
microscopic understanding of unconventional superconductivity. An
important first step towards this daunting task is the
identification of the correct phenomenological description of the
relevant materials. Indeed, our current understanding of the cuprate
\cite{James_adv_phys}, heavy fermion \cite{Sigrist&Ueda}, ruthenate
\cite{Sigrist&Rice}, colbaltate \cite{Johannes&Mazin}, and
quasi-two-dimensional organic \cite{group2d} superconductors depends
on phenomenological descriptions as in each of these cases there is
no widely agreed upon microscopic description of the
superconductivity. However, despite long standing evidence
\cite{PFdisorder,ClOdisorder} of unconventional superconductivity in
the Bechgaard salts and theoretical proposals of triplet states
\cite{Abrikosov}, the correct phenomenological description of the
superconducting state has not, until now, been identified.

In this Letter, we perform a group theoretical classification of
\emph{all} of the possible superconducting states in the Bechgaard
salts that respect translational symmetry. This shows that there are
only eight symmetry distinct states. By considering the properties
of these states we show that only one of them is consistent with the
full range of thermodynamic measurements
\cite{
Oh,LeeNMRa,LeeNMRb,LukeClO,Lee97,Lee00,HS-ClO,PF-activated,ClO-activated}
that have been performed on both \PF and \ClOn. This state is
somewhat analogous to the polar state, first discussed in the
context of $^3$He.

\emph{Symmetry analysis}: The Bechgaard salts form triclinic
crystals whose symmetry is represented by the $C_i$ point group.
$C_i$ contains only two elements, the identity and inversion. Thus
the point group only differentiates between even and odd parity
states (which we henceforth refer to as $s$-wave and $p$-wave states
respectively). Note that symmetry does not distinguish, say
`$d$-wave' states from $s$-wave states or `$f$-wave' states from
$p$-wave states as the crystal has neither rotational nor mirror
symmetries.  (Hence the terms `$d$-wave' and `$f$-wave' are rather
meaningless in the context of the Bechgaard salts
.) All the superconducting states unambiguously identified in nature
thus far, are even under frequency reversal. However, this is not
required \emph{a priori} in the superconducting state and so we must
distinguish between even- or odd-frequency pairing \cite{odd-freq}.
As the wavefunction of a fermionic system must be antisymmetric
under the exchange of \emph{all} labels, the allowed states are
then: even-frequency, $s$-wave singlet; odd-frequency, $p$-wave,
singlet; even-frequency, $p$-wave triplet; and odd-frequency,
$s$-wave triplet.

The gap function of the singlet phases may be written as
$\Delta({\bf k})=\eta\psi_{\bf k}$, where $\psi_{\bf k}$ may be any
function with the appropriate parity that satisfies translation
invariance, and $\eta$ is the complex Ginzburg-Landau (GL) order
parameter. Thus there are only two symmetry distinct singlet states:
the conventional $s$-wave, even-frequency singlet and a $p$-wave,
odd-frequency singlet.


To describe triplet superconductivity one must introduce a complex
vector gap function, ${\bf d}({\bf k})$ \cite{Vollhardt}. The
interpretation of ${\bf d}({\bf k})$ is straightforward (at least
for unitary vectors \cite{unitary}) as it points along the $S_z=0$
projection, i.e., perpendicular to the spin of the Cooper pairs. For
triplet superconductors we must distinguish between weak and strong
spin-orbit coupling (SOC). If the SOC is sufficiently weak we may
rotate the spin and spatial degrees of freedom independently;
therefore the symmetry group of the normal state is ${\cal
G}=SO(3)\otimes G\otimes U(1)\otimes\cal T$, where $SO(3)$ is spin
rotation symmetry, $G$ is the point group of the crystal ($C_i$ for
the Bechgaard salts), $U(1)$ is gauge symmetry and $\cal T$ is time
reversal symmetry. However, for strong SOC the independent rotation
of the spin and spatial degrees of freedom is not a symmetry of the
system. Therefore, the symmetry group becomes ${\cal
G}=G^{(J)}\otimes U(1)\otimes\cal T$, where the group $G^{(J)}$ is
identical to the usual point group of the crystal except that the
operations of the point group simultaneously act on both the spin
and the spatial degrees of freedom.

In general when SOC is strong we expect that ${\bf d}({\bf
k})\propto({\bf \hat{a}}X_{\bf k},{\bf \hat{b}'}Y_{\bf k},{\bf
\hat{c}^*}Z_{\bf k})$, where $X_{\bf k}$, $Y_{\bf k}$, and $Z_{\bf
k}$ are arbitrary functions which respect translational symmetry and
transform like $k_x$, $k_y$, and $k_z$ respectively under the
symmetry operations of the point group
\cite{James_adv_phys,Sigrist&Ueda}. However, as $C_i$ contains only
the identity and inversion the only restriction on $X_{\bf k}$,
$Y_{\bf k}$, and $Z_{\bf k}$ is that they have the appropriate
parity. Thus, ${\bf d}({\bf k})=\eta({\bf \hat{a}}\psi_{\bf
k}^{a},{\bf \hat{b}'}\psi_{\bf k}^{b},{\bf \hat{c}^*}\psi_{\bf
k}^{c})$, where $\psi_{\bf k}^{a}$, $\psi_{\bf k}^{b}$ and
$\psi_{\bf k}^{c}$ may be any functions which have the required
parity and satisfy translation symmetry, and the GL order parameter
is a single complex number ($\eta$). Therefore the GL free energy is
$F_s-F_n=\alpha|\eta|^2 + \beta|\eta|^4$.
Clearly there is only one solution and therefore there is only one
symmetry distinct spin part of the wavefunction for triplet
superconductivity when SOC is strong. These states are analogous to
the BW phase, which is realised at ambient pressure in $^3$He, but
may be either an even-frequency $p$-wave or an odd-frequency
$s$-wave state.

Because $C_i$ only contains one-dimensional irreducible
representations ${\bf d}({\bf k})=\vec\eta\cdot{\bf\Psi}_{\bf k}$
for weak SOC, where $\vec\eta$ is the complex vector GL order
parameter and ${\bf\Psi}_{\bf k}=(\psi_{\bf k}^{a}, \psi_{\bf
k}^{b}, \psi_{\bf k}^{c})$. The GL expression for the free energy is
thus \cite{James_adv_phys,Sigrist&Ueda}
$F_s-F_n=\alpha|\vec\eta|^2 + \beta_1|\vec\eta|^4 +
\beta_2\left|\vec\eta\cdot\vec\eta\right|^2$. 
The ground state, up to arbitrary rotations in spin-space, is
$\vec\eta\propto(1,i,0)$ for $\beta_2>0$ and
$\vec\eta\propto(1,0,0)$ for $\beta_2<0$. We refer to these states
as the $\beta$ and polar phases respectively by analogy with $^3$He
\cite{Vollhardt}. The $\beta$ phase corresponds to pairing in a
single spin channel, while that polar phase corresponds to pairing
in both equal spin pairing (ESP) channels
. Note that a representation must be at least two-dimensional for
the ABM phase (which is realised under pressure in $^3$He
\cite{Vollhardt} and probably in \SRO \cite{Sigrist&Rice}) to be
possible and so this phase can be immediately ruled out of our
consideration of the Bechgaard salts. Both the $\beta$ and polar
states may exist as either $p$-wave, even-frequency, triplet states
or as $s$-wave, odd-frequency, triplet states. Thus there are four
possible triplet states if SOC is weak.

\begin{table*}
\begin{tabular}{ccc|cccccccc}
  \hline
  state & $\begin{tabular}{c} spin-\\orbit\\coupling \end{tabular}$ & ${\bf d}({\bf k})$ &
  $\begin{tabular}{c} (symmetry\\required)\\nodes \end{tabular}$ &
  BTRS & $\begin{tabular}{c} Pauli\\limited\\${\bf H}\|{\bf a}$ \end{tabular}$ &
  $\begin{tabular}{c} Pauli\\limited\\${\bf H}\|{\bf b}'$ \end{tabular}$ &
  $\begin{tabular}{c} $\underline{\chi_s(0)}$\\${\chi_n}$\\${\bf H}\|{\bf a}$ \end{tabular}$ &
  $\begin{tabular}{c} $\underline{\chi_s(0)}$\\${\chi_n}$\\${\bf H}\|{\bf b}'$ \end{tabular}$ &
  $\begin{tabular}{c} disorder\\suppresses\\$T_c$\end{tabular}$ &
  $\begin{tabular}{c} Hebel-\\Slichter\\peak\end{tabular}$  \\
  \hline
  s-singlet & any & - & {\bf no} & {\bf no} & yes & yes & 0 & 0 & no & yes \\
  polar s-triplet & weak & $(1,0,0)$ & gapless & {\bf no} & {\bf no} & {\bf no} & {\bf 1} & {\bf 1} & {\bf yes} & {\bf no} \\
  $\beta$ s-triplet & weak & $(1,i,0)$ & gapless & yes & {\bf no} & {\bf no} & {\bf 1} & {\bf 1} & {\bf yes} & {\bf no} \\
  BW s-triplet & strong & $({\bf \hat{a}}\psi_{g\bf k}^{a},{\bf \hat{b}'}\psi_{g\bf k}^{b},{\bf \hat{c}^*}\psi_{g\bf k}^{c})$
        & gapless & {\bf no} & {\bf no} & {\bf no} & 2/3$^a$ & 2/3$^a$ & {\bf yes} & {\bf no} \\
  p-singlet & any & - & gapless & {\bf no} & yes & yes & 0 & 0 & {\bf yes} & {\bf no} \\
  {\bf polar p-triplet} & {\bf weak} & {$\bf(1,0,0)$} & {\bf no} & {\bf no} & {\bf no} & {\bf no} & {\bf 1} & {\bf 1} & {\bf yes} & {\bf no} \\
  $\beta$ p-triplet & weak & $(1,i,0)$ & {\bf no} & yes & {\bf no} & {\bf no} & {\bf 1} & {\bf 1} & {\bf yes} & {\bf no} \\
  BW p-triplet & strong & $({\bf \hat{a}}\psi_{u\bf k}^{a},{\bf \hat{b}'}\psi_{u\bf k}^{b},{\bf \hat{c}^*}\psi_{u\bf k}^{c})$
        & {\bf no} & {\bf no} & {\bf no} & {\bf no} & 2/3$^a$ & 2/3$^a$ & {\bf yes} & {\bf no} \\
  \hline
  \PF & & & no\footnote{See main text for discussion and caveats.} \cite{PF-activated} & ? & no \cite{Lee97} & no \cite{Lee97} & 1 \cite{LeeNMRa} & 1 \cite{LeeNMRb} & yes \cite{PFdisorder} & no$^a$  \cite{LeeNMRa} \\
  \ClO & & & no \cite{ClO-activated} & no \cite{LukeClO} & ? & no \cite{Oh} & ? & ? & yes \cite{ClOdisorder} & no \cite{HS-ClO} \\
  \hline
\end{tabular}
\caption{Summary of the thermodynamic properties of the eight
symmetry distinct states allowed for superconductors with $C_i$
point groups and comparison with experiments on the Bechgaard salts.
We see that only the polar $p$-wave triplet state is compatible with
experiment.
$\chi_s(0)/\chi_n$ is the ratio of the spin susceptibility in the
limit $T\rightarrow0$ to that in the normal state above $T_c$. The
basis functions $\psi_{u\bf k}^{i}$ ($\psi_{g\bf k}^{i}$) may be any
odd (even) parity functions which satisfy translation symmetry. The
symbol `?' in the experimental sections indicates that an experiment
has not been performed.}\label{tab:summary}
\end{table*}

\emph{Properties of the superconducting states}: We list all eight
symmetry distinct superconducting states for the group $C_i$, which
represents the symmetry of crystals of the Bechgaard salts, in table
\ref{tab:summary}. Our task is now to determine the properties of
the of these states and to compare these properties with those found
experimentally in \PF and \ClOn.

None of the four even-frequency states are required by symmetry to
have nodes in the order parameter. This extremely unusual property
for an unconventional superconductor is \emph{not}, as is often
stated, because of the quasi-one-dimensionality of the Fermi
surface, but is a direct consequence of the extremely low symmetry
of the Bechgaard salts. Recall that the basis functions may be any
function with the appropriate parity. Therefore, $s$-wave states
have no symmetry required nodes and $p$-wave states are required to
vanish only at the origin ($\Gamma$-point), which, by symmetry, the
Fermi surface may not cross. In contrast, odd-frequency pairing
states are intrinsically gapless \cite{odd-freq}. Specific heat
measurements on \PF \cite{PF-activated} and thermal conductivity
measurement on \ClO \cite{ClO-activated} both show an exponentially
activated behaviour, suggestive of a nodeless gap. As these
experiments see a full gap they are inconsistent with odd-frequency
pairing. However, the NMR relaxation rate, $1/T_1$, has a power law
temperature dependence \cite{LeeNMRa,LeeNMRb}. If this power law
were assumed to arise from quasiparticles then it would be
suggestive of nodes in the gap. Rostunov \etal \cite{Rostunov} have
recently shown that, in a triplet superconductor, collective
spin-wave excitations can also lead to a power law dependence of
$1/T_1$. This theory may also resolve the puzzle of why the power
law dependence of $1/T_1$ is seen even at temperatures very close to
the critical temperature \cite{LeeNMRa,LeeNMRb}, which is not
expected from nodal quasiparticles.  Thus these experiments may
suggest a triplet pairing state.

A extremely small peak is seen in $1/T_1$, just below $T_c$
\cite{LeeNMRa,HS-ClO}. However, this peak more than an order of
magnitude smaller than the Hebel-Slichter expected for an
even-frequency, $s$-wave, singlet order parameter \cite{TinkhamSC}.
This strongly suggests that the even-frequency, $s$-wave, singlet
order parameter is \emph{not} realised in the Bechgaard salts.
Further evidence for this conclusion comes from the observed strong
suppression of the superconducting critical temperature by disorder
\cite{PFdisorder,ClOdisorder}. The only state for which this
suppression of $T_c$ by disorder is \emph{not} expected is the
even-frequency, $s$-wave, singlet order parameter \cite{disorder}.

Evidence for triplet pairing comes from the observation that the
upper critical fields  in the conducting planes of \PF and \ClO
exceed the weak coupling Clogston-Chandrasekhar (or Pauli) limit by
more than a factor of four \cite{Lee97,Lee00,Oh}. The Pauli limit
occurs when the Zeeman energy penalty for forming $S_z=0$ Cooper
pairs exceeds the condensation energy gained by entering the
superconducting state and applies to singlet states and pairs in the
$S_z=0$ projection of a triplet state \cite{PAG}. When SOC is weak
the spin part of the order parameter is not `pinned' to the lattice.
Therefore, the superconductor may minimise its energy by aligning
${\bf d}({\bf k})\perp\bf H$ \cite{PAG,foot-perp}. Thus, the triplet
phases for weak SOC will always be ESP phases in the reference frame
of the magnetic field and are therefore not Pauli limited.

In contrast the triplet phases for strong SOC are `pinned' to the
lattice as the symmetry group does not allow the independent
rotation of the spin and spatial degrees of freedom. When a field
which exceeds that Pauli limit is applied to system, it will
completely suppress the pairing in the $S_z=0$ channel, i.e., ${\bf
d}({\bf k})$ goes to zero in the direction parallel to the field.
However, ESP is not suppressed, i.e., ${\bf d}({\bf k})$ remains
finite perpendicular to the field. Both possible triplet states for
strong SOC in the Bechgaard salts have finite components
perpendicular to the conducting plane and thus we do not expect them
to be Pauli limited. Hence, we do not expect any of the symmetry
distinct triplet states to be Pauli limited. Therefore, while the
large critical field is strong evidence against singlet pairing it
does not differentiate among the six candidate triplet states. It is
also worth noting that calculations suggest that the observed
critical field is too large to be accounted for by FFLO singlet
states which break translational symmetry \cite{Lebed}. Such states
are therefore not considered in detail in this Letter.

If a superconducting order parameter breaks TRS then spontaneous
supercurrents will flow around impurities and near grain boundaries
\cite{
Sigrist}. The most sensitive probe of these tiny currents is the
zero field muon spin relaxation (ZF-$\mu$SR) rate. Small fields
consistent with broken TRS (BTRS) have been observed in 
UPt$_3$ \cite{LukeUPt3}, 
U$_{1-x}$Th$_x$Be$_{13}$ \cite{Heffner}, 
PrOs$_4$Sb$_{12}$ \cite{Aoki},
and 
\SRO \cite{Luke}. Importantly, as the currents due to a
superconducting order parameter with BTRS are extremely small the
magnetic fields they generate can be suppressed by very small
longitudinal fields (50~G is sufficient in \SRO \cite{Luke}). 
Luke \etal \cite{LukeClO} measured the ZF-$\mu$SR rate in \ClO and
did not find any increase in the relaxation rate to within their
experimental resolution ($\approx25$~G). This indicates that the
superconducting state of \ClO does not break TRS. Both the even- and
odd-frequency pairing $\beta$ phases are inconsistent with this
experiment.


The spin susceptibility, $\chi_s(T)$ strongly distinguishes between
different triplet states \cite{Vollhardt}. When an $S_z=0$ pair
forms it no longer contributes to the spin susceptibility and thus
for a singlet superconductor or a triplet superconductor with ${\bf
d}({\bf k})\|\bf H$, $\chi_s(T)\rightarrow0$ as $T\rightarrow0$. On
the other hand ESP does not affect the susceptibility. Thus,
$\chi_s(T)$ does not change upon passing though $T_c$ for ESP-only
states such as the ABM phase of $^3$He \cite{Vollhardt}. For triplet
states which contain both ESP and $S_z=0$ pairs the decrease in
$\chi_s(T)$ is proportional to the fraction of $S_z=0$ pairs. As
discussed above, the symmetry distinct states for weak SOC are all
ESP states and thus we expect $\chi_s(T)/\chi_n=1$ for all $T<T_c$,
where $\chi_n$ is the spin susceptibility of the normal state. In
contrast, in both the possible states for strong SOC contain pairing
in all three $S_z$ projections. Thus,
\begin{eqnarray}
\frac{\chi_s^i(0)}{\chi_n} \rightarrow 1-\frac{\langle| \psi_{\bf
k}^{i}|^2 \rangle_{FS}}{\langle |{\bf\Psi}_{\bf k}|^2 \rangle_{FS}},
\label{eqn:Kinght}
\end{eqnarray}
where, $\langle\dots\rangle_{FS}$ indicates the average over the
Fermi surface, $i\in\{a,b,c\}$, and the superscript on the
susceptibility indicates the orientation of the field
\cite{Vollhardt}. If the averages over the three basis functions are
the same, as they are in the BW phase of $^3$He,
$\chi_s(0)/\chi_n=2/3$ for all field orientations. No decrease in
$\chi(T)$ is detected below $T_c$ in \PF with the field aligned
along either the $\bf a$ \cite{LeeNMRa} or $\bf b'$ \cite{LeeNMRb}
axes. $\chi_s^a(0)/\chi_n=\chi_s^b(0)/\chi_n=1$ if and only if
$\psi_{\bf k}^{a}$ and $\psi_{\bf k}^{b}$ vanish everywhere on the
Fermi surface. As we do not, in general, expect $\psi_{\bf k}^{a}$
and $\psi_{\bf k}^{b}$ to vanish everywhere on the Fermi surface the
BW state is incompatible with the measured Knight shift.

We summarise the properties of the eight symmetry distinct
superconducting phases in table \ref{tab:summary}. It can readily be
seen that the only state consistent with all of the experiments is
the `polar' $p$-wave triplet state realised for weak SOC which is
specified by ${\bf d}({\bf k})\propto(\psi_{\bf k},0,0)$. It is
worth stressing the very small number of assumptions that have been
made to reach this conclusion: (i) translational symmetry is not
violated by superconducting state of the Bechgaard salts; (ii) there
is not an accidental vanishing of two independent basis functions if
SOC is strong; and (iii) that the superconducting states of \PF and
\ClO have the same symmetry, i.e., that the materials are related by
`chemical pressure'. (i) is supported by calculations of the
critical field for the FFLO state in these materials \cite{Lebed},
but a triplet analogue of FFLO cannot formally be ruled out at this
stage. However, it is difficult to see how such a phase would be
stabilised. (ii) can be tested experimentally:
$\chi_s(T)\rightarrow0$ as $T\rightarrow0$ with ${\bf H}\|\bf c^*$
for the BW state with $\langle| \psi_{\bf k}^{a}|^2
\rangle_{FS}=\langle| \psi_{\bf k}^{b}|^2 \rangle_{FS}=0$; whereas
for the the polar state $\chi_s(T)=\chi_n$ for ${\bf H}\|\bf c^*$.
(iii) is easily tested experimentally; in particular measurements of
the Knight shift in \ClO and the ZF-$\mu$SR rate in \PF would
complete the set of measurements required to uniquely determine the
superconducting state of each material individually. For
sufficiently small magnetic fields SOC will be strong therefore our
analysis predicts that a Fredericks transition
\cite{Vollhardt,bjpthesis} from the BW state to the polar state
occurs at extremely low fields. The different states correspond to
different broken symmetries therefore measurements of collective
modes \cite{Vollhardt,Rostunov} could also provide confirmation of
our identification of the superconducting state.

\emph{Previous theoretical work}: Early theoretical work only
focused on whether the superconductivity was singlet or triplet and
did not propose a specific triplet state \cite{Abrikosov}. 
The state discussed by Lebed and coworkers \cite{Lebed} assumes
strong SOC and is therefore a special case of the BW phase. Lebed
\etaln's state has $\langle |\psi_{\bf k}^{b}|^2 \rangle_{FS} =0$,
but $\langle |\psi_{\bf k}^a|^2 \rangle_{FS} \ne0$. Eq.
(\ref{eqn:Kinght}) shows that this theory will predict a large
decrease in the spin susceptibility below $T_c$ when ${\bf H}\|{\bf
a}$. This prediction is clearly contradicted by experiment
\cite{LeeNMRa} and therefore this theory can be ruled out. Duncan
\etal \cite{Duncan} have discussed the symmetry distinct triplet
states in an orthorhombic ($D_{2h}$) crystal. As the Bechgaard salts
are triclinic and the angles involved are rather large this is
\emph{not} a good approximation. Nevertheless the state they propose
[a `$p_x$' state, ${\bf d}({\bf k})\propto(k_x,0,0)$ and weak SOC]
is a special case of the polar state we have shown to be the actual
superconducting state. However, the `$p_x$' state has an accidental
node in the plane $k_x=0$. As this plane does not cut the Fermi
surface the physical properties of the `$p_x$' state are rather
similar to polar state. However, we stress that the node in the
`$p_x$' state is \emph{not} required by symmetry and will raise the
energy of the state, therefore this node is unphysical. Shimahara
\cite{Shimahara} proposed that a singlet state is found at low field
and a triplet state is found in large fields. As inversion is a
symmetry of the crystal, such a change must be accompanied by a
phase transition. This has not been observed and so this theory does
not seem compatible with experiment.


%

\emph{In conclusion} we have shown that their are eight symmetry
distinct superconducting states in monoclinic crystals with the
$C_i$ point group. Of these only the $p$-wave, even-frequency, polar
state [${\bf d}({\bf k})\propto(\psi_{u{\bf k}},0,0)$] is consistent
with the full range of experiments on both \ClO and \PFn. There is
not yet sufficient experimental evidence to determine the
superconducting state of either these materials individually, but
the chemical pressure hypothesis suggests that the polar state is
also realised in these materials. Finally, we note that the same
symmetry analysis applies to the Fabre salts.

\acknowledgements

This work was motivated by conversations with  A. Ardavan. I thank
J. Annett, 
S. Blundell, J. Fj{\ae}restad,  R. McKenzie, F. Pratt, and J.
Varghese for enlightening conversations. This work was funded by the
ARC.


\begin{thebibliography}{999}
\bibitem{foot-perp}
An intrinsically complex vector, such as the order parameter of the
$\beta$ phase, $(1,i,0)$, is defined as being perpendicular to a
field if ${\bf d}({\bf k})\cdot{\bf H}=0$.
\end{thebibliography}
\end{document}